\documentclass{elsart}

\usepackage{graphicx,amssymb}

\newcommand{\calF}{{\mathcal F}}
\newcommand{\calT}{{\mathcal T}}
\newcommand{\calA}{{\mathcal A}}
\newcommand{\calB}{{\mathcal B}}

\newcommand{\initial}{{| {f_x}^{(0)}, {f'_y}^{(0)} \rangle} }

\newcommand{\fxinit}{{{f_x}^{(0)}}}
\newcommand{\fyinit}{{{f'_y}^{(0)}}}

\newcommand{\inita}{|a_x^{(0)}, a_y^{(0)} \rangle}

\begin{document}
\begin{frontmatter}
\title{
Topological degeneracy of non-Abelian states
{\sc for dummies}}

\author[ISSP]{Masaki Oshikawa}
\ead{oshikawa@issp.u-tokyo.ac.jp}
\author[Toronto,KIAS]{Yong Baek Kim}
\author[UCR]{Kirill Shtengel}
\author[ProjQ,UCLA]{Chetan Nayak}
\author[UMD]{Sumanta Tewari}

\address[ISSP]{Institute for Solid State Physics,
University of Tokyo, Kashiwa 277-8581 Japan}
\address[Toronto]{Department of Physics, University of Toronto,
Toronto ON M5S 1A7, Canada}
\address[KIAS]{School of Physics, Korea Institute for Advanced Study,
Seoul 130-722, Korea}
\address[UCR]{Department of Physics,
University of California, Riverside, CA 92521, USA}
\address[ProjQ]{Microsoft Research, Project Q, Kohn Hall,
University of California, Santa Barbara, CA 93106-4030, USA}
\address[UCLA]{Department of Physics and Astronomy,
University of California, Los Angeles, CA 90095-1547, USA}
\address[UMD]{Condensed Matter Theory Center, Department of Physics,
University of Maryland, College Park, MD 20742, USA}

\begin{abstract}
We present a physical construction of degenerate
groundstates of the Moore-Read Pfaffian states,
which exhibits non-Abelian statistics,
on general Riemann surface with genus $g$.
The construction is given by a
generalization of the recent argument
[M.O. and T. Senthil, Phys. Rev. Lett. {\bf 96}, 060601 (2006)]
which relates fractionalization and topological order.
The nontrivial groundstate degeneracy obtained
by Read and Green [Phys. Rev. B {\bf 61}, 10267 (2000)]
based on differential geometry is reproduced exactly.
Some restrictions on the statistics, due to the
fractional charge of the quasiparticle are also discussed.
Furthermore, the groundstate degeneracy
of the $p+ip$ superconductor in two dimensions,
which is closely related to the Pfaffian states,
is discussed with a similar construction.
\end{abstract}

\end{frontmatter}

\section{Introduction}

Many quantum phases and phase transitions can be understood
with a local order parameter.
However, it has been noticed that there is a class
of distinct quantum phases which cannot be characterized by
any local order parameter.
The underlying `order' of these phases is dubbed as
`topological order' \cite{Wen90}.
One of the signatures of topological order, which has
proven to be particularly useful in numerical studies of
quantum Hall systems, is
the dependence of the ground state degeneracy on the topology
of the manifold on which the system is defined.
Such a groundstate degeneracy is called `topological
degeneracy' to distinguish it from
the case of ordinary spontaneous symmetry breaking
with a local order parameter, where the groundstate degeneracy
is determined by the pattern of the symmetry breaking and
thus does not depend on the topology.
In addition to its conceptual interest, there is a renewed
interest in topological degeneracy due to its potential
for the realization of qubits, as topologically
degenerate groundstates are expected to be stable against
external perturbations coupling to local observables \cite{Kitaev97}.

Particle statistics is a fascinating
aspect of quantum mechanics.
In general dimensions, the quantum state of identical
particles should be either unchanged or receive the overall
phase factor $(-1)$ under an exchange of a pair of particles.
These correspond to Bose and Fermi statistics, respectively.
However, in 2 dimensions, there are more possibilities.
Namely, the quantum state can acquire a nontrivial phase
factor $e^{i\theta}$, where the real parameter $\theta$
is called statistical angle, under the exchange.
This phenomenon is called anyonic statistics (for $\theta \neq 0, \pi$).
Such a statistics is indeed realized for quasiparticles/holes
in fractional quantum Hall liquids.

Anyonic statistics can be further generalized to
the situation in which multi-quasiparticle states
(which can be viewed as ground states on the
multi-punctured sphere) are topologically degenerate.
Then one of these ground states can transform
into another under the exchange of identical particles.
Such a transformation can be parameterized by a matrix.
This phenomenon is called non-Abelian statistics,
as the matrices corresponding to different exchanges
generally do not commute.
Again, the non-Abelian statistics can be realized,
at least theoretically, in several
proposed fractional quantum Hall states.
The first and most discussed example is the
Pfaffian state constructed by Moore and Read \cite{Moore91}.
Numerical exact diagonalization studies \cite{Morf98}
of small systems indicate that this state is realized at the observed
$\sigma_{xy}=\frac{5}{2}\,\frac{e^2}{h}$ fractional quantum Hall
plateau \cite{Xia04}.
Several experiments have been proposed which could confirm this
\cite{Fradkin98,DasSarma05,Bonderson06,Stern06}.
It has also been suggested that a very similar state (differing
only in the Abelian part of quasiparticle braiding statistics)
is realized in Sr$_2$RuO$_4$ \cite{DasSarma06a}
and ultra-cold fermions with a
$p$-wave Feschbach resonance in atomic traps \cite{GRA,TDNZZ}.
The non-Abelian statistics of quasiparticles in this state has formed the basis
of proposals \cite{DasSarma05,Bravyi05} for topological quantum
computation \cite{DasSarma06b}.
The unitary transformations resulting from particle exchanges
are virtually errorless since they depend only
on the topological class of the particles'
trajectories\cite{Kitaev97}.

Quantum number fractionalization is another
intriguing concept in the quantum many-body problem:
the charge carried by a quasiparticle, which
is an elementary excitation of the strongly-correlated
ground state, can be fractional with respect to that
of the original constituent particle (electron).
While the notion of fractionalization might seem to be
quite independent of the two others introduced above,
recently, a direct and close connection among these three
concepts was demonstrated \cite{OshikawaSenthil}.
(See also Refs.~\cite{WHK,SKW}.)
Namely, in a gapped system, topological degeneracy
follows from the assumption that the quasiparticle
carries a fractional charge:
when the charge of the quasiparticle is $p/q$ in the unit of
the electron charge, the groundstate degeneracy $N_g$
on the two-dimensional surface with genus $g$ is shown to satisfy
\begin{equation}
  N_g \geq q^g.
\label{eq:minNg}
\end{equation}
Moreover, the (minimal) topological degeneracy in such a case is
also affected by the quasiparticle statistics.
For example, if the quasiparticles obey either Fermi or Bose
statistics, eq.~(\ref{eq:minNg}) can be replaced by the
stronger condition
\begin{equation}
  N_g \geq q^{2g}.
\label{eq:minNg2}
\end{equation}

However, in Ref.~\cite{OshikawaSenthil}, the case of
non-Abelian statistics is not studied in detail.
The general lower bound~(\ref{eq:minNg}) was derived without
relying on any assumption about the statistics, and thus is
expected to hold for both the Abelian and non-Abelian cases.
Nevertheless, it would be worthwhile to study the topological
degeneracy in systems with non-Abelian statistics from this
new perspective.

In this paper, we discuss the topological groundstate degeneracy
of the Moore-Read Pfaffian state, which exhibits
non-Abelian statistics, by generalizing the arguments in
Ref.~\cite{OshikawaSenthil}. The topological degeneracy
of the Pfaffian state is rather nontrivial.
The degeneracy on a two-dimensional surface of
genus $g$ has been related by Read and Green \cite{Read00}
to ``spin structures'' in differential geometry.
In this paper, we will show an alternative, more ``physical''
derivation of the topological degeneracy
(which might be actually related to
the original argument based on an index theorem).
The groundstate degeneracy of weak-pairing $p+ip$ superconductor
in two dimensions
is also discussed based on a similar construction.
In the course of doing this, we also make
several observations which would be
also valid for other systems with non-Abelian statistics.
In particular, we show that the Abelian part of
the statistics of a quasiparticle can be restricted
by its effective charge.

\section{Topological degeneracy from fractionalization}
\label{sec:revOS}

Here we give a brief review of Ref.~\cite{OshikawaSenthil}
to make the paper self-contained.

Let us assume that we have a two-dimensional system
of interacting electrons of charge $e$.
We further assume that there is a finite gap above the
groundstate(s), and that the elementary excitations are
quasiparticles/holes with the charge $e^*=(p/q)e$, where
$p$ and $q$ are coprimes.
First, let us consider the two-dimensional torus.
We introduce the adiabatic insertion $\calF_x$ of the unit
flux quantum in the ``hole'' of the torus,
so that a vector potential is induced in the $x$-direction.
Furthermore, we define the process $\calT_x$,
in which a quasiparticle-quasihole pair is created and then
the quasiparticle is dragged along the $x$-direction to wrap around
the system before it is pair-annihilated with the quasihole.
The Aharonov-Bohm effect for this fractionally charged
quasiparticle implies
\begin{equation}
 \calT_x \calF_x
= \calF_x \calT_x e^{2\pi i e^*/e}
= \calF_x \calT_x e^{2\pi i p/q} .
\label{eq:magx}
\end{equation}
(To make this more precise, the adiabatic flux insertion $\calF_x$
must be accompanied by a large gauge transformation.
For simplicity we do not make it explicit in this paper,
although it is essential for the argument to work.
The details can be found in Ref.~\cite{OshikawaSenthil}.)
It is natural to expect that both of these processes map
a groundstate to a groundstate, which we assume for now.
The ``magnetic algebra'' (\ref{eq:magx}) acting on the groundstate
subspace immediately implies that there must be at least
$q$ groundstates on the torus.

The same construction can be done in the
$y$-direction, resulting in the magnetic algebra
\begin{equation}
 \calT_y \calF_y = \calF_y \calT_y e^{2\pi i p/q} .
\label{eq:magy}
\end{equation}
It appears that the two copies of the algebra lead to
a $q^2$-fold degeneracy on the torus.
However, this is not always the case because the two algebras
may not be actually independent, depending on the
quasiparticle statistics.
For the case of Abelian statistics, we have
\begin{equation}
 \calT_x \calT_y = \calT_y \calT_x e^{2 i \theta},
\label{eq:TxTy}
\end{equation}
where $\theta$ is the statistical angle.
For either Bose ($\theta=0$) or Fermi ($\theta=\pi$) statistics,
$\calT_x$ and $\calT_y$ commute.
In this case, the minimum degeneracy on the torus is indeed $q^2$-fold.
On the other hand, as we have noted above,
for other statistics, the two magnetic algebras are not independent,
so the degeneracy does not have to be $q^2$-fold.
In fact, the Laughlin state at the filling factor $\nu=1/q$
exhibits the fractionalization with charge $(1/q)e$ but
the groundstate degeneracy on the torus is only $q$-fold.
This can be understood as follows.
In the Laughlin state, a quasiparticle/hole can be identified
with a vortex/antivortex, namely the flux tube with unit flux
quantum piercing through the two-dimensional surface.
Thus, as pointed out by Wen and Niu~\cite{Wen90},
the process $\calT_x$ actually introduces
a unit flux quantum through the ``hole'' of the torus, just as
$\calF_y$ does.
Therefore, as far as
the action on the groundstate subspace is concerned,
we can identify $\calT_x$ and $\calF_y$,
although they are not completely identical.
Similarly, $\calT_y$ can be identified with ${\calF_x}^{-1}$.
Thus the two copies of the magnetic algebra
are actually reduced to one for the Laughlin state.

\begin{figure}
\includegraphics[width=0.8\columnwidth]{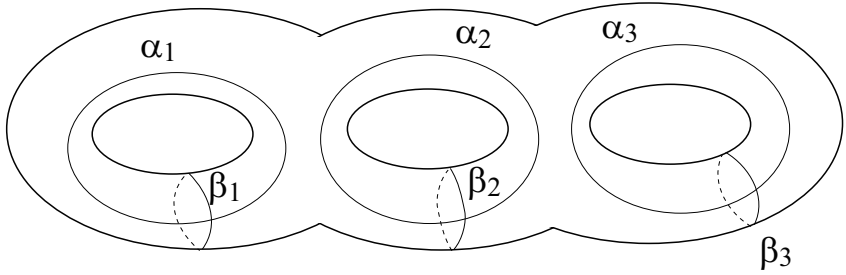}
\caption{
Riemann surface and its elementary cycles.
The figure shows the example with $g=3$, with the three pairs
of intersecting elementary cycles $(\alpha_j, \beta_j)$,
$j=1,2,3$, each of which is associated to a ``hole''.
Elementary cycles belong to different pairs do not intersect
with each other.
}
\label{fig:Riemann}
\end{figure}

The above consideration can be readily generalized to
the general two-dimensional surface with genus $g$.
As shown in Fig.~\ref{fig:Riemann},
for genus $g$, there are $g$ pairs of intersecting elementary cycles
$(\alpha_1, \beta_1), (\alpha_2, \beta_2), \ldots, (\alpha_g, \beta_g)$.
For each cycle $\gamma = \alpha_j$ or $\gamma=\beta_j$,
we can construct the generalization of eq.~(\ref{eq:magx}).
Among them, $g$ magnetic algebras (one taken from each pair)
are always independent of each other.
For example we can take the magnetic algebras of all $\alpha_j$'s,
which are independent.
Thus the minimum degeneracy of $q^g$ follows.
In particular, in the Laughlin state
the magnetic algebra for $\beta_j$ cycle reduces to that
of the intersecting $\alpha_j$ cycle, as in the case of $y$ and $x$
of the torus.
Thus the groundstate degeneracy derived from the present analysis
is only $q^g$, which is indeed the correct degeneracy for the
Laughlin state.

On the other hand, for either Bose or Fermi statistics,
$\calT_{\alpha_j}$ and $\calT_{\beta_j}$ commute.
This leads to the minimum degeneracy $q^{2g}$.

\section{The Pfaffian state and its non-Abelian statistics}
\label{sec:Pfaff}

In this section, we summarize the minimal knowledge
of the Pfaffian state and its non-Abelian statistics
needed for the next section.

The Pfaffian state proposed by Moore and Read~\cite{Moore91} is
characterized by the following wave function of particles
in a magnetic field (on an infinite plane)
\begin{equation}
 \Psi_{\mathrm{Pf}}(z_1,z_2, \ldots, z_N)
 = {\mathrm{Pf}}{\left( \frac{1}{z_i - z_j} \right)}
 \prod_{i<j} (z_i - z_j)^{M+1} \exp{(-\frac{1}{4}\sum_j |z_j|^2)} ,
\end{equation}
where Pf denotes the Pfaffian, which is the square
root of the determinant of an antisymmetric matrix
\begin{equation}
{\mathrm{Pf}}{\left( \frac{1}{z_i - z_j} \right)} =
{\calA}\left\{ \frac{1}{z_1 - z_2} \,\frac{1}{z_3 - z_4}
\,\frac{1}{z_5 - z_6} \,\ldots\right\}
\end{equation}
(${\calA}$ is the antisymmetrization symbol).
$z_j$ is the complex coordinate of the $j$-th fundamental particle.
$M$ is a non-negative integer characterizing the Pfaffian state.
For $M$ even, the fundamental particle is a boson
while it is a fermion for $M$ odd. The latter is the case
of relevance to the quantum Hall effect. We emphasize that
this refers to the statistics of the fundamental
particle and is not to be confused with the statistics of the
quasiparticles that are elementary excitations.
In this paper, we call the fundamental, constituent particle
an `electron', even though it is a boson for $M$ even.
The Landau level filling fraction for this state is:
\begin{equation}
\nu = \frac{1}{M+1} .
\end{equation}
$M=1$ corresponds to (fermionic) electron state with $\nu=1/2$.

The Pfaffian state has quasiparticles and quasiholes with
fractional charge and non-Abelian statistics.~\cite{Moore91,Nayak96}
Initially, the Pfaffian state was discussed in relation to conformal field
theory and a Chern-Simons effective field theory.
Later, the nature of the Pfaffian state was elucidated with
the BCS-type pairing picture.~\cite{Read00}
In this paper, we follow the latter approach which is more
accessible to a wider audience.

In the pairing picture,
the Pfaffian state can be understood as a ``weak-pairing''
$p$-wave ($p+ip$) paired state~\cite{Read00} of `composite fermions'.
The composite fermion is an electron
with $M+1$ unit flux quanta attached.
The attachment of the flux quanta transforms the statistics
of the composite particle to fermionic from the `bosonic' electron
if $M$ is even, and keeps the statistics fermionic if $M$ is odd.
In either case, the composite particle is a fermion
and thus is called as the composite fermion (CF).~\cite{CFbook}
The CF has the same charge $e$ as the electron does.
As in the Laughlin state,
a quasiparticle/hole can be identified with a vortex/antivortex.
However, as the CFs are now paired, the minimal
flux tube that can be contained in a vortex is now
a half flux quantum, not a unit flux quantum.
The calculation of the quasiparticle charge can be done
similarly to the case of the Laughlin state, and we obtain
\begin{equation}
e^* = \frac{1}{2} \nu e = \frac{1}{2(M+1)} e,
\label{eq:eofPf}
\end{equation}
where the factor $1/2$ comes from the fact that
the vortex contains a half flux quantum.

Remarkably, at each vortex (or antivortex), there is
a Majorana (real) fermion bound state,
which is topologically stable. The Majorana fermions
associated with a pair of vortices can be combined
into a Dirac fermion zero mode, which can be occupied
or unoccupied. This (non-local) two-level system
gives the degeneracy required for non-Abelian statistics.

We note that, although the ``real'' $p+ip$ superconductor
consisting of pairs of electrons (not CFs) is closely related
to the Pfaffian states, it is different from any of the
Pfaffian states labelled by the integer $M$.
A vortex in the
$p+ip$ superconductor carries a Majorana fermion boundstate,
however, as an important difference,
does {\em not} a fractional charge as in eq.~(\ref{eq:eofPf}).
We will discuss the topological degeneracy of
the $p+ip$ superconductor later in Sec.~\ref{sec:p+ip}.

Ivanov~\cite{Ivanov01} explicitly showed how non-Abelian statistics
can be derived directly from the existence of the Majorana
fermion boundstate at a vortex.
Let us consider a system with $2n$ vortices (quasiparticles).
As we commented above, each vortex has a Majorana fermion
bound state, which we shall denote as $\eta_j$ ($j=1,2,\ldots, 2n$).
They are characterized by the anticommutation relation
\begin{equation}
 \{ \eta_j, \eta_k \} = 2 \delta_{jk} .
\label{eq:Majorana}
\end{equation}
These Majorana (real) fermions can be combined into $n$ complex
fermions as
\begin{equation}
  c_j \equiv \frac{1}{2} (\eta_{2j-1} + i \eta_{2j}),
  \;\;\;\;
  c^{\dagger}_j \equiv \frac{1}{2} (\eta_{2j-1} - i \eta_{2j}),
\label{eq:psidef}
\end{equation}
so that $c_j$ satisfies the standard anticommutation relation
\begin{eqnarray}
\{ c_j, c_k \} = \{ c^{\dagger}_j, c^{\dagger}_k \} &=& 0,
\\
\{ c^{\dagger}_j, c_k \} &=& \delta_{jk} .
\end{eqnarray}
The Hilbert space of these bound state
fermions are $2^n$-dimensional,
corresponding to the eigenvalue of $c^{\dagger}_j c_j = 0$ or
$1$ for each $j$. We will refer to these two possibilities
by saying that in the former case the two quasiparticles
`fuse to the vacuum' while in the latter case they
`fuse to a fermion'.
This gives the representation space of the
braid group, leading to the non-Abelian statistics.
In fact, since the parity of the fermion number is conserved,
the Hilbert space can be split into even and odd fermion number
sectors. Each sector is of dimension $2^{n-1}$, and gives an irreducible
representation of the Braid group.~\cite{Nayak96}
In this paper, we disregard this reduction and
just work with the full $2^n$-dimensional Hilbert space.

Ivanov~\cite{Ivanov01} pointed out that the Majorana
fermion changes sign
when the phase of the superconducting order parameter changes
by $2\pi$. This immediately leads to the representation
of the elementary exchange $T_j$ of two vortices $j$ and $j+1$:
\begin{equation}
T_j: \left\{
\begin{array}{ll}
\eta_j \rightarrow \eta_{j+1}, \\
\eta_{j+1} \rightarrow - \eta_j, \\
\eta_k \rightarrow \eta_k & ( k \neq j, j+1)
\end{array}
\right.
\end{equation}
This encapsulates the non-Abelian statistics of quasiparticles
and quasiholes. To make this more explicit, we define
the representation of the exchange $T_j$ on
the zero-mode Hilbert space:
\begin{equation}
\label{eq:T-def}
 \tau(T_j) = e^{i\theta} \exp{\big( \frac{\pi}{4} \eta_{j+1} \eta_j \big)},
\end{equation}
where $\theta$ is the (Abelian) statistical angle which is
an arbitrary real parameter at this point. (In the quantum Hall
context of interest to us, $\theta$ will depend on the filling
fraction.)
In terms of the complex fermion~(\ref{eq:psidef}),
\begin{equation}
 \tau(T_{2j-1}) = e^{i\theta}
  \exp{\big[ i \frac{\pi}{4} (2 c_j^{\dagger} c_j -1 ) \big]}
  = e^{i\theta}
  \exp{\big( i \frac{\pi}{4} \sigma_z^{(j)} \big)},
\label{eq:Tsigma}
\end{equation}
where we have defined $\sigma_z^{(j)}=2 c_j^{\dagger} c_j -1$.
$\sigma_z^{(j)}$ has eigenvalues $\pm 1$.

The topological degeneracy of the Pfaffian state is
rather nontrivial and there was apparently
some confusion in the early days.
The correct result was given by
Read and Green~\cite{Read00}, who
related the topological degeneracy
to the ``spin structure'' of differential geometry.
According to them, the degeneracy depends
on whether the electron number (or equivalently, the number
of CFs) is even or odd.
For an even electron number, the groundstate degeneracy is
\begin{equation}
 N_g = (M+1)^g \, 2^{g-1} (2^g +1),
\label{eq:evenNg}
\end{equation}
on the two-dimensional surface with genus $g$.
For an odd electron number, it reads
\begin{equation}
 N_g = (M+1)^g \, 2^{g-1} (2^g - 1),
\label{eq:oddNg}
\end{equation}
Here, the factor $2^{g-1}(2^g \pm 1)$ is the number of
spin structures, and $(M+1)^g$ comes from the U(1) charge
sector.

Let us make a few observations about these degeneracies
in relation to the general argument~\cite{OshikawaSenthil}
reviewed in Sec.~\ref{sec:revOS}.
In the present case, the minimum degeneracy
eq.~(\ref{eq:minNg}) (which is independent of the statistics)
is $q^g = 2^g(M+1)^g$.
This lower bound is indeed satisfied in the even electron number case,
but not in the odd electron number case at $g=1$.
On the other hand, according to the argument in Ref.~\cite{OshikawaSenthil},
it appears to possible to generate $q^g$ multiplets of
groundstates starting from a groundstate.
Thus one might expect that the groundstate
degeneracy would be an integral multiple of the minimum
degeneracy $q^g = 2^g (M+1)^g$.
However, this is not the case in either the even or
odd electron number sectors.
In this paper, we present a resolution to this puzzle,
with an explicit derivation of the groundstate degeneracy.

\section{Consistency between non-Abelian statistics and
charge fractionalization}
\label{sec:consistency}

We now generalize the argument in Ref.~\cite{OshikawaSenthil}
to the Pfaffian state.

Our starting point is the identification of the quasiparticle/hole
as a vortex/antivortex, as we reviewed in Sec.~\ref{sec:Pfaff}.
Because the vortex carries only a half flux quantum,
the process $\calT_y$ introduces just half a flux quantum
into the ``hole'' of the torus.
Thus $\calT_y$ cannot be identified with $\calF_x$ in the Pfaffian
state, in contrast to the Laughlin state.
Rather, we can postulate
\begin{equation}
 \calF_x \sim {\calT_y}^2,
\label{eq:FxTy2}
\end{equation}
in the Pfaffian state, as doing the operation
$\calT_y$ twice introduces
a unit flux quantum just as $\calF_x$ does.

The commutation relation between $\calT_x$ and $\calT_y$
can be related to the statistics of quasiparticles,
as was discussed in Ref.~\cite{Wen90} on the Abelian case.
A naive generalization of the Abelian case~(\ref{eq:TxTy})
to the Pfaffian case using the representation eq.~(\ref{eq:T-def})
would read
\begin{equation}
 \calT_x \calT_y = \calT_y \calT_x \,e^{2 i \theta}
  \:i{\eta_3}{\eta_1}.
\label{eq:TxTysig}
\end{equation}
Here, we have labeled the Majorana
fermion operators associated to the two quasiparticles
$\eta_1$, $\eta_3$; those associated to the two quasiholes
are $\eta_2$, $\eta_4$. When one quasiparticle
is taken around the other, the unitary transformation
$e^{2 i \theta}  \:{\eta_3}{\eta_1}$ is applied to the
state of two quasiparticle-quasihole pairs.
By assumption, the quasiparticle-quasihole
pair used to define $\calT_x$ fuses to the vacuum (since
they were created out of the vacuum), i.e.
$i{\eta_1}{\eta_2}=1$, and similarly for the
quasiparticle-quasihole pair used to define
$\calT_y$, i.e. $i{\eta_3}{\eta_4}=1$. However,
the operation $\calT_x \calT_y  \calT^{-1}_x \calT^{-1}_y$
is not proportional to the identity (or even diagonal) in the
basis of eigenstates of $i{\eta_1}{\eta_2}$, $i{\eta_3}{\eta_4}$.
Therefore, an operator appears on the right-hand side
of (\ref{eq:TxTysig}). On the other hand, $\calT_x$, $\calT_y$
are supposed to be operators which transform one
ground state into another, so operators acting
on a multi-quasiparticle Hilbert space cannot appear
on the right-hand side of (\ref{eq:TxTysig}).
Therefore, this equation isn't quite right.
However, it is still a useful heuristic
as we will discuss in the following.

Now let us discuss the commutation relation between
$\calF_x$ and $\calT_x$.
The fractional charge of the quasiparticle
implies~\cite{OshikawaSenthil} that
\begin{equation}
 \calF_x \calT_x = \calT_x \calF_x e^{-2\pi i e^*/e}
= \calT_x \calF_x e^{-i \pi /(M+1)} .
\label{eq:FxTxC}
\end{equation}
On the other hand, because of eq.~(\ref{eq:FxTy2}), the
commutation relation is identical to that
of ${\calT_y}^2$  and ${\calT_x}$.

The ansatz~(\ref{eq:TxTysig}) based on eq.~(\ref{eq:Tsigma}) implies
\begin{equation}
\calF_x \calT_x = \calT_x \calF_x e^{-4 i \theta}
  {\left(i{\eta_3}{\eta_1}\right)^2}
= \calT_x \calF_x e^{- 4i\theta}  .
\label{eq:FxTxB}
\end{equation}
The second equality follows because
${\left(i{\eta_3}{\eta_1}\right)^2}=1$.
Now, remarkably, the commutation relation is Abelian, namely
the factor $e^{-4i\theta}$ is just a phase factor.
The commutation relation~(\ref{eq:FxTxB})
is related to four consecutive exchanges of a pair
of quasiparticles, as the commutation relation
of $\calT_x$ and $\calT_y$ is related to two consecutive exchanges.
In fact, Ivanov noticed that four consecutive exchanges
reduces to a Abelian phase factor~\cite{Ivanov01}.
Our analysis indicates that this is indeed required by
consistency with the ``charge'' relation~(\ref{eq:FxTxC}).

Furthermore, the consistency between eqs.~(\ref{eq:FxTxB})
and~(\ref{eq:FxTxC}) actually determines the Abelian part
of the statistics, which is not easy to obtain from the
approach taken in Ref.~\cite{Ivanov01}.
Consistency between (\ref{eq:FxTxB})
and (\ref{eq:FxTxC}) requires
\begin{equation}
  e^{-4i\theta} = e^{-\pi i / (M+1)}.
\end{equation}
Namely,
\begin{equation}
  \theta = \frac{\pi}{4(M+1)} + \frac{\pi}{2} n ,
\end{equation}
where $n$ is an integer.
This includes the correct value $\theta = {\pi}/{4(M+1)}$.
To see that this is the correct value, we note that
the Abelian part of the statistics is the same
in both the `weak pairing' and `strong pairing' limits;
only in the former limit is there a non-Abelian part as well.
In the `strong pairing' limit, the $\nu=1/(M+1)$ quantum
Hall state is a $\nu=1/[4(M+1)]$ state of charge $2e$ pairs.
The quasiparticles have charge ${e^*} = (2e)/[4(M+1)]=e/[2(M+1)]$
and statistics $\theta=\pi/[4(M+1)]$.

\section{Groundstate degeneracy on a torus}

\subsection{Construction of the groundstates}

In order to study the groundstate degeneracy of the Pfaffian
state on a torus, we classify the groundstates
according to the eigenvalues of simultaneously
diagonalizable operators.
A natural candidate would be the set $\calF_x$,
$\calF_y$.
Actually, generally they do not
commute with the Hamiltonian, as the flux insertion
accelerates the electrons by the induced electric field
and thus changes the energy.~\cite{Oshikawa03}
Nevertheless, by our assumption that they map a groundstate to
a groundstate, $\calF$ operators applied on a groundstate
does not change the energy (up to an exponentially small
finite-size correction).
Thus, the $\calF$ operator should be diagonalizable in
the groundstate subspace, so that the groundstates
are classified by its eigenvalue.

However, our ansatz $\calF_x \sim \calT_y^2$ and
the charge relation~(\ref{eq:FxTxC}) in the $y$ direction
($\calF_y \calT_y = \calT_y \calF_y e^{ - i \pi / (M+1)}$)
imply
\begin{equation}
\calF_x \calF_y \sim \calF_y \calF_x e^{ 2 \pi i / (M+1)} .
\end{equation}
Thus, except for $M=0$, $\calF_x$ and $\calF_y$
do not commute and hence cannot be diagonalized simultaneously.
Instead, as a set of commuting operators,
we can take $\calF_x$ and ${\calF_y}^{M+1}$.
Let us label a groundstate as
\begin{equation}
 | f_x, f'_y \rangle,
\end{equation}
where $f_x$ and $f'_y$ are the eigenvalues of $\calF_x$
and ${\calF_y}^{M+1}$, respectively.

The commutation relation ~(\ref{eq:FxTxC}) and
$\calF_y \sim {\calT_x}^{-2}$ implies
\begin{equation}
 \calT_x | f_x, f'_y \rangle \sim | e^{- i \pi / (M+1)} f_x, f'_y \rangle ,
\end{equation}
where we have disregarded the phase of the wavefunction,
which is not important in what follows.
Similarly, we obtain for $\calT_y$
\begin{equation}
 \calT_y | f_x, f'_y \rangle \sim | f_x,  - f'_y \rangle ,
\end{equation}
where we have used $\big(e^{-i \pi /(M+1)}\big)^{M+1} = -1$.

We start from an ``initial'' groundstate
\begin{equation}
|\Psi_0 \rangle = \initial
\end{equation}
and construct other groundstates by applying $\calT_{x,y}$.

Then it seems that we can construct $4(M+1)$ groundstates
(including the original one $|\Psi_0\rangle$),
as follows.
\begin{equation}
\begin{array}{cc}
|\Psi_0\rangle = \initial  & \calT_y | \Psi_0 \rangle \sim | \fxinit, - \fyinit \rangle \\
\calT_x | \Psi_0 \rangle \sim | e^{-i \pi/(M+1)} \fxinit, \fyinit \rangle
& \calT_y \calT_x | \Psi_0 \rangle \sim | e^{-i\pi/(M+1)} \fxinit , - \fyinit \rangle \\
{\calT_x}^2 | \Psi_0 \rangle \sim | e^{-i \frac{2\pi}{(M+1)}} \fxinit, \fyinit \rangle
& \calT_y {\calT_x}^2 | \Psi_0 \rangle
\sim | e^{- i \frac{2\pi}{M+1}} \fxinit , - \fyinit \rangle \\
\vdots & \vdots \\
{\calT_x}^{2M+1} | \Psi_0 \rangle \sim | e^{-i \frac{(2M+1) \pi}{M+1}} \fxinit, \fyinit \rangle
& \calT_y {\calT_x}^{2M+1} | \Psi_0 \rangle
\sim | e^{-i \frac{(2M+1) \pi}{M+1}} \fxinit, - \fyinit \rangle
\end{array}
\label{eq:4M+4states}
\end{equation}
These states should be orthogonal to each other because they can
be distinguished by different eigenvalues of $\calF_x$
and ${\calF_y}^{M+1}$.

In particular, for the $M=0$ bosonic Pfaffian state at $\nu=1$
and the $M=1$ fermionic Pfaffian state at $\nu=1/2$,
the above logic would give ground state degeneracies
of $4$ and $8$, respectively.

However, according to Read and Green~\cite{Read00},
the correct groundstate degeneracy on the
torus is $3 (M+1)$ and $M+1$
in the even and odd electron number sectors, respectively.
In Ref.~\cite{Read00} it was also argued that,
on the torus, the odd electron number sector corresponds
to the boundary condition $++$ (periodic
boundary conditions in both directions).
The even electron number sector corresponds to the other
three boundary conditions $+-$,$-+$, and $--$,
which have an antiperiodic boundary condition
along at least one direction.

The $4(M+1)$ groundstates we have obtained above
are too many for either the even or odd electron number
cases, although it is indeed the sum of the degeneracies
in both sectors.

\subsection{Blocking mechanism}

The resolution to the problem of ``too many groundstates''
is actually deeply connected to the non-Abelian nature
of the Pfaffian state.

The preceding discussion is based on the assumption that
application of $\calT_{x,y}$ to a groundstate gives
a groundstate. Namely, after the pair annihilation,
the system should go back into the vacuum (groundstate).
This would be true in the case of Abelian statistics,
where there is a single fusion channel for
a quasiparticle and quasihole.
However, in the case of non-Abelian statistics,
there are multiple fusion channels (in this case, two)
for a quasiparticle and quasihole. One of these
fusion channels is the vacuum, but the other isn't,
so the pair might leave behind an excitation rather
than annihilate to the ground state.
Here we reexamine the state $\calT_x | \Psi_0 \rangle$,
with a careful consideration of this.

For simplicity, let us first consider periodic boundary
conditions along both directions, i.e. $++$ boundary conditions.
A quasiparticle-hole (vortex-antivortex) pair is created
on the initial vacuum $| \Psi_0 \rangle$.
Let us denote the Majorana fermion boundstate carried by
the quasihole and the quasiparticle by $\eta_1$ and $\eta_2$,
respectively. Just after the pair creation,
the internal state is given by $|0\rangle$ which satisfies
\begin{equation}
 c |0\rangle = 0,
\end{equation}
for the complex fermion $c \equiv (\eta_1 + i \eta_2)/2$.
The pair is created out of the ground state, so there
cannot be an extra fermion.
Similarly, the internal state of the pair
{\em must be in the same state $|0\rangle$ before the
pair can be annihilated.} Otherwise, there will be
a fermion left over.

We point out that before the pair is annihilated,
the relative locations of the pair must be the same as
they were immediately after the pair was created.
This is because the quasiparticle/hole are identified with
a vortex/antivortex, which induce a phase change for
CF operators as they are taken around them.
As the vortex/antivortex in the Pfaffian state
contains a half unit flux quantum, the CF operator
changes its phase by $\pi$ around the vortex/antivortex.
Thus, a string of the ``branch cut'' should be attached to
the vortex/antivortex (quasiparticle/hole) as
emphasized by Ivanov~\cite{Ivanov01}.
The phase of the CF operator jumps by $\pi$
when passing the branch cut.

\begin{figure}
\centerline{\includegraphics[height=2in]{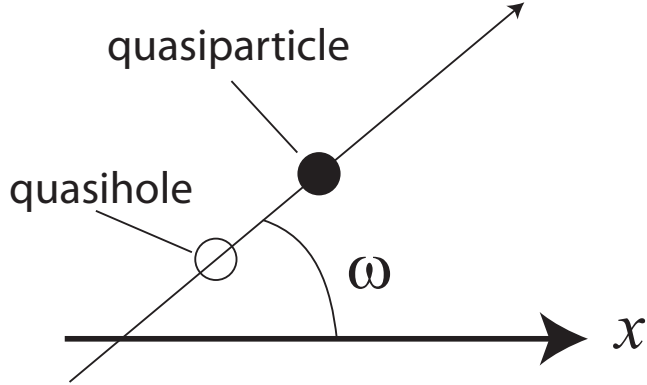}}
\caption{The relative location of the quasiparticle and quasihole
is characterized by the angle $\omega$ between the $x$ axis
and the position vector of the quasiparticle relative to the
quasihole.}
\label{fig:qpqhlocation}
\end{figure}

Let the angle $\omega$ be the relative
direction of the quasiparticle/hole with respect to
the $x$ axis, as defined in Fig.~\ref{fig:qpqhlocation}.
Suppose that $\omega=0$ just after pair creation.
Then, as $\omega$ changes, the phase of the CF
operators at the quasiparticle and quasihole change as
\begin{equation}
  \psi_{CF} \rightarrow \psi_{CF} e^{- i \omega/2},
\;\;\;\;\;
  \psi_{CF} \rightarrow \psi_{CF} e^{i \omega/2},
\end{equation}
respectively.
Since the Majorana fermion operator of the boundstate
is given in terms of the CF operator as
$ \eta \sim u \psi_{CF} + v \psi^{\dagger}_{CF}$,
the Majorana fermion operator is also changed
from the original one.

Thus the pair cannot be annihilated,
unless $\omega$ is an integral multiple of $2\pi$
and the Majorana fermions return to their original form.
In fact, when $\omega$ is an odd integral multiple
of $2\pi$, the CF operator and, thus, the Majorana
fermion change sign: $\psi \rightarrow - \psi,
\eta \rightarrow - \eta$.
However, because both of the Majorana fermions
$\eta_1$ and $\eta_2$ change sign,
the complex fermion operator just changes sign:
$c \rightarrow - c$.
The pair annihilation is still possible since
$-c$ does annihilate $|0\rangle$.

Let us now turn to the reexamination of the
state $\calT_x | \Psi_0 \rangle$.
After pair creation,
the created quasiparticle is dragged around
the $x$ cycle of the torus.
Then we would like to pair annihilate it with the quasihole.
Now we argue that {\it one} of the Majorana fermions has
its sign flipped, due to a subtle statistical effect.

\begin{figure}
\centerline{\includegraphics[height=3in]{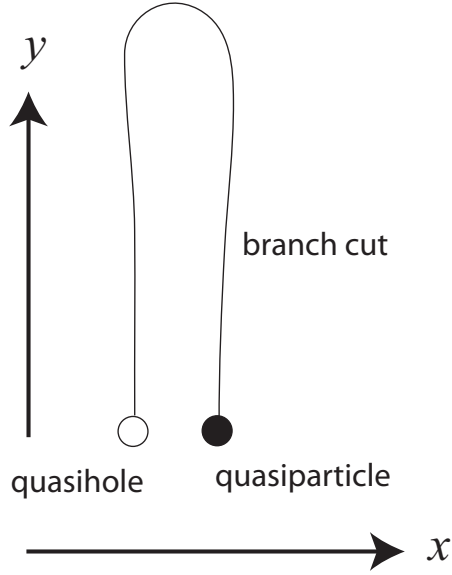}}
\caption{Initial configuration of quasiparticle/hole,
just after the pair was created.
For convenience, we extend the branch cut to the $y$ direction,
so that the branch cuts meet at a distant point.}
\label{fig:qpqhinitial}
\end{figure}

\begin{figure}
\centerline{\includegraphics[height=2in]{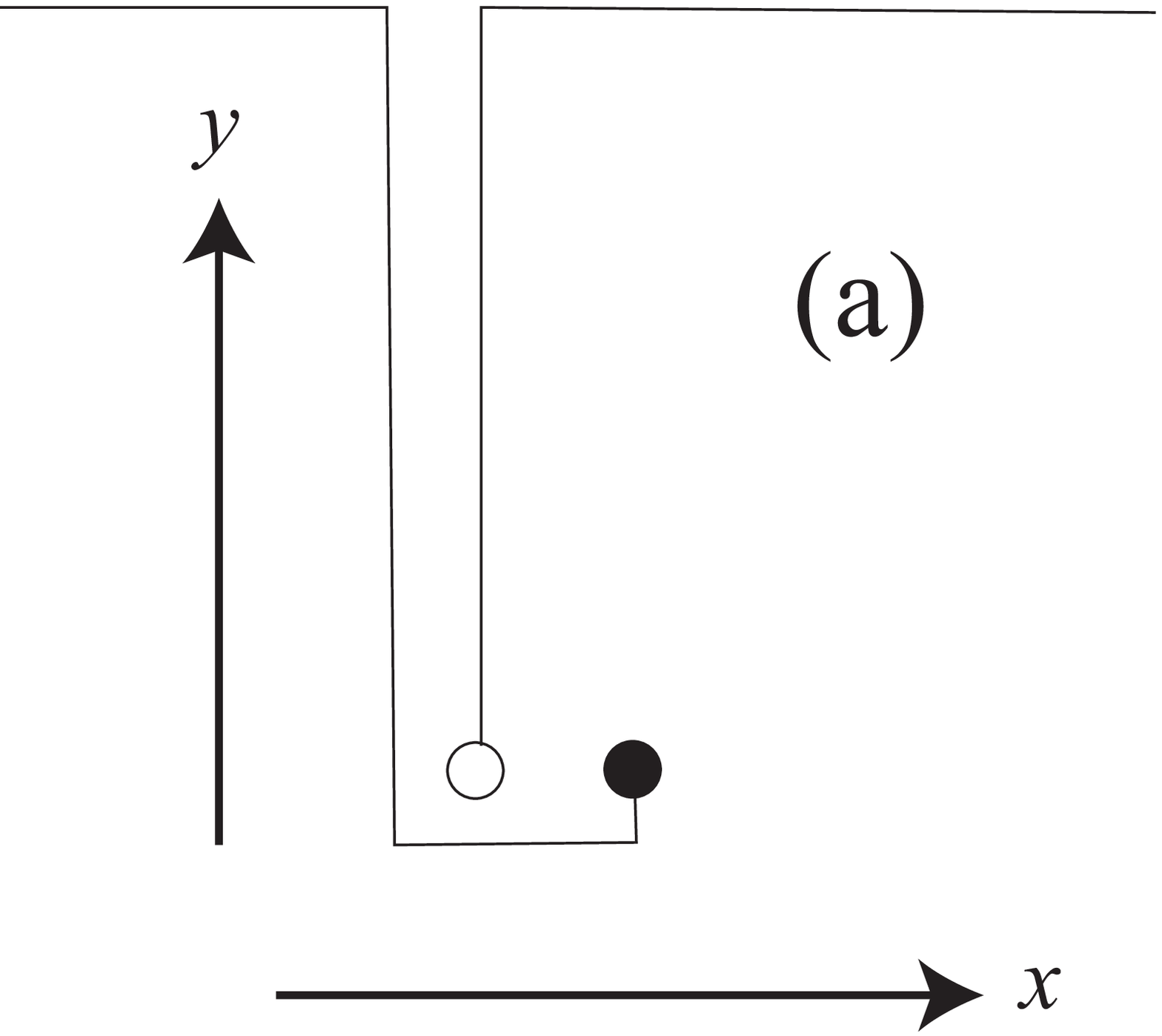}{\hskip 2.5 cm}
\includegraphics[height=2in]{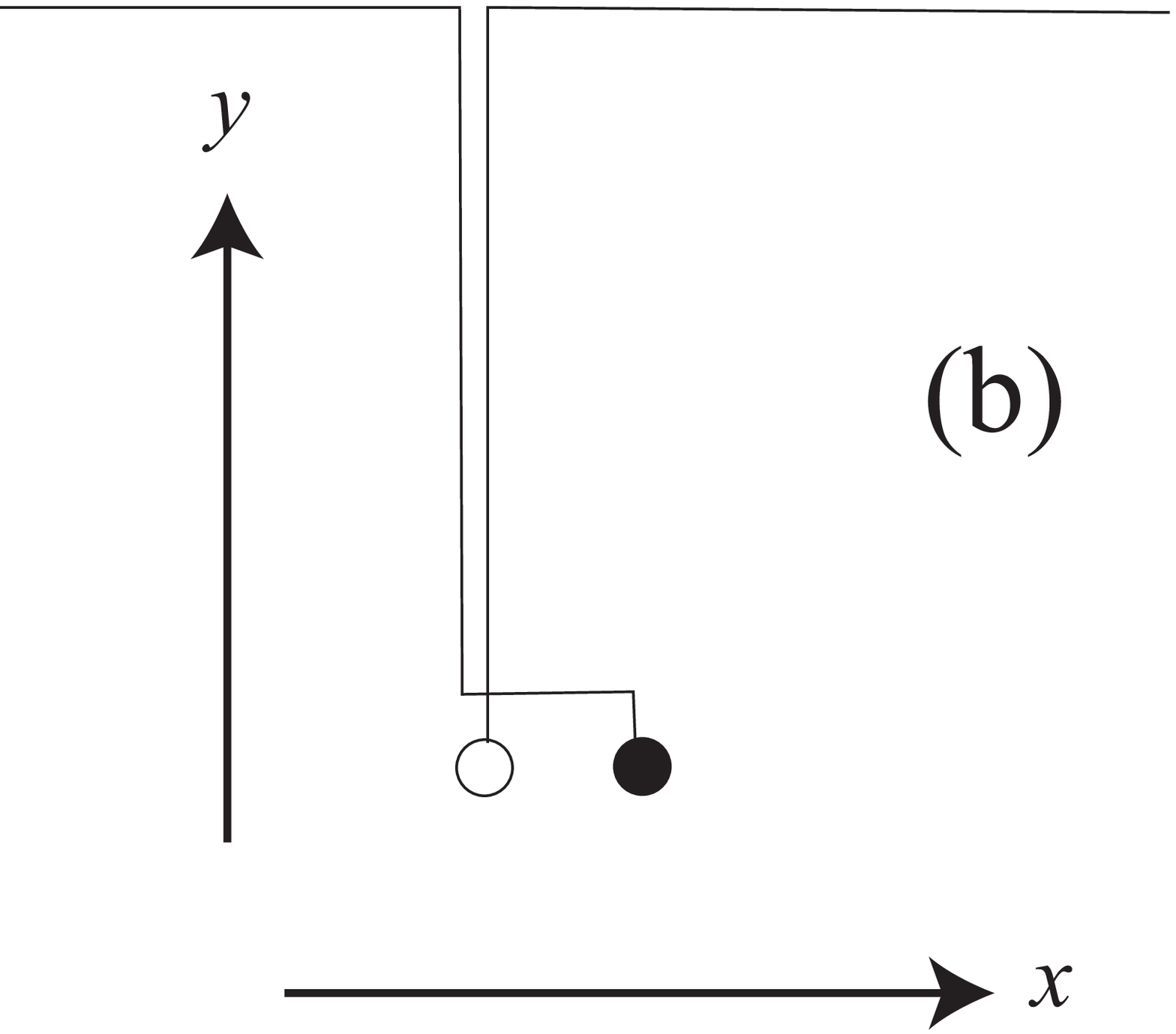}}
\caption{The final configuration of quasiparticle/hole,
after the quasiparticle has encircled the torus in the $x$ direction.
On the left (a), the quasiparticle came back to its original place,
moving below the quasihole, while on the right (b), it
travelled above the quasihole.}
\label{fig:qpqhfinal1}
\end{figure}

Let the original relative locations of the pair
just after pair creation be $\omega=0$.
For convenience, we extend the branch cut attached
to the quasiparticle/hole (vortex/antivortex)
in the $y$ direction (we may assume that they are connected
at a distant point.)
The initial configuration with the branch cut
is shown in Fig.~\ref{fig:qpqhinitial}.
The quasiparticle is then dragged in the $x$ direction
to encircle the torus.
As discussed above, the pair must return to the original
relative locations, before being annihilated.
Depending on the particular path taken by the quasiparticle,
the final configuration would look like
Fig.~\ref{fig:qpqhfinal1}(a) and (b).

In either case, a branch cut parallel to $x$ direction
is introduced by the application of $\calT_x$.
This corresponds to a change of the boundary condition
along $y$ direction to the antiperiodic one.
This is also naturally understood as $\calT_x$
introduces a half unit flux quantum into
the hole of the torus.

\begin{figure}[t]
\centerline{\includegraphics[height=2.5in]{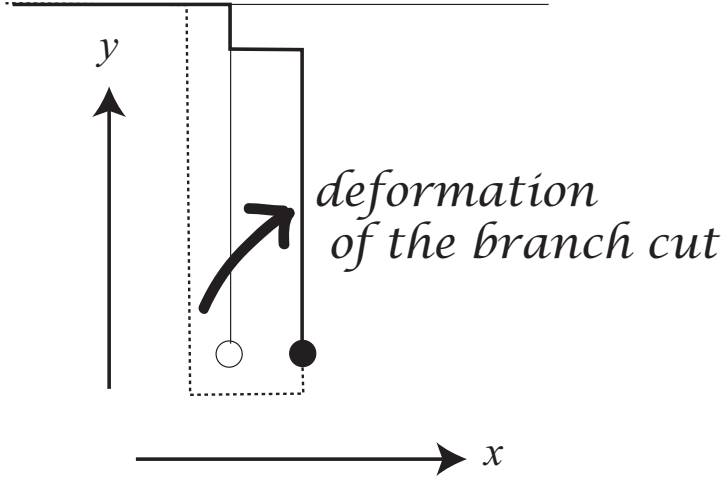}}
\caption{Deformation of the branch cut from
Fig.~\protect\ref{fig:qpqhfinal1}(a).
The configuration returns to the
original one in Fig.~\ref{fig:qpqhinitial},
except for the branch cut running parallel to
$x$ direction at a distance. This corresponds to
the change of the boundary condition along $y$ direction.
During the deformation, the branch cut passes through
the quasihole, flipping the sign of the Majorana fermion
associated with the quasihole.
}
\label{fig:qpqhdeform1}
\end{figure}

There is another effect which is important for the
present discussion.
In the former scenario, Fig.~\ref{fig:qpqhfinal1}(a),
the phase of CF operators
of the quasihole has changed by $\pi$.
This can be understood by moving the branch cut
as in Fig.~\ref{fig:qpqhdeform1} so that it matches
the original configuration,
in a similar manner to Ivanov's derivation of the
statistics.~\cite{Ivanov01}
In the process, the branch cut passes through the
quasihole, making the phase change of $\pi$ on
the quasihole.
This induces flipping of the Majorana fermion~\cite{Ivanov01}
$\eta_2 \rightarrow - \eta_2$.

If we take the other scenario, Fig.~\ref{fig:qpqhfinal1}(b),
the branch cut does not pass through the quasihole
even in the deformation process similar to Fig.~\ref{fig:qpqhdeform1}.
However, in this case, the quasiparticle had already crossed
the branch cut before returning into the original location.
Thus the Majorana fermion associated with the quasiparticle
is flipped in its sign: $\eta_1 \rightarrow - \eta_1$.

As a consequence, the complex fermion operator is transformed
according to:
\begin{equation}
  c \rightarrow \pm c^{\dagger},
\label{eq:psi2dag}
\end{equation}
where the sign is positive and negative, respectively,
for the two cases in Fig.~\ref{fig:qpqhfinal1}(a) and (b).
Either way, the complex fermion operator
does not annihilate the quantum state $|0\rangle$.
Thus pair annihilation to the vacuum does not
occur.

This description in terms of the evolution of the
Majorana fermion operators corresponds to the ``Heisenberg picture''.
Alternatively, we can discuss the boundstates
in the ``Schr\"odinger'' picture, as follows.
In the former scenario (Fig.~\ref{fig:qpqhfinal1}(a)),
the appropriate time evolution operator
(on the internal state of the quasiparticle/hole)
in the Schr\"odinger picture
is given by $U = e^{i\varphi} \eta_1$, where $\varphi$
is an arbitrary real phase factor.
This can be verified by
\begin{eqnarray}
  U^{-1} \eta_1 U &=& \eta_1 ,  \\
  U^{-1} \eta_2 U &=& - \eta_2,
\end{eqnarray}
which are derived from the fundamental anticommutation relation
eq.~(\ref{eq:Majorana}).
Thus, in the Schr\"odinger picture, the internal state
after the quasiparticle winding process is given by
\begin{equation}
 U | 0 \rangle = e^{i\varphi} (c + c^\dagger) |0 \rangle
 = e^{i \varphi} | 1 \rangle,
\end{equation}
where $|1 \rangle = \psi^{\dagger}|0\rangle$.
Since this is orthogonal to the initial internal state
which appears after the pair creation, the quasiparticle-hole
pair cannot be annihilated.
In the latter scenario, Fig.~\ref{fig:qpqhfinal1}(b),
a similar result follows by taking
$U= e^{i\varphi} \eta_2$.

Thus we conclude that
the purported groundstate $\calT_x | \Psi_0 \rangle$
actually does not exist, as the quasiparticle/hole pair
is blocked from pair annihilation owing to the
effect of their non-Abelian statistics.
Similarly, $\calT_y | \Psi_0 \rangle$ and
$\calT_x \calT_y | \Psi_0 \rangle$
also do not exist as groundstates.
(Actually, the mechanism of
the absence of the groundstate $\calT_x \calT_y | \Psi_0 \rangle$
is not quite trivial, as we will discuss later.)

On the other hand,
${\calT_x}^2$ may be interpreted
as creation of a quasiparticle/hole pair, and
dragging the quasiparticle so that it encircles {\em twice},
and then annihilation of the pair.
By encircling the torus twice, the sign of the
Majorana fermion remains the same.
Thus the pair can be annihilated, and
${\calT_x}^2 | \Psi_0 \rangle$ does give a groundstate.
We can apply ${\calT_x}^2$ repeatedly, up to $M$ times,
to produce different eigenvalues of $\calF_x$.
In the present construction,
therefore we obtain $M+1$ orthogonal groundstates
\begin{equation}
 {\calT_x}^{2n} | \Psi_0 \rangle ,
\end{equation}
for $n=0,1,2,\ldots,M$.
Since the ${\calT_x}^2$ does not change the boundary
conditions, all these states belong to the
boundary condition $++$.
Indeed this is consistent with Ref.~\cite{Read00},
in which only $M+1$ groundstates are predicted for
the periodic boundary conditions $++$.

\subsection{Twisting the block}

Now let us consider other boundary conditions.
For simplicity, we start from the antiperiodic-antiperiodic
one $--$.
This is equivalent to having half unit flux quantum trapped in
each hole of the torus.
When a quasiparticle encircles the torus,
it is affected by the flux due to the Aharonov--Bohm effect.
The CF operator at the quasiparticle
is flipped in its sign due to the antiperiodic boundary condition,
thereby inducing a sign flip of the Majorana fermion
associated with the quasiparticle.
(This should not be confused with the fractional phase
$e^{2\pi i e^*/e}$ acquired by the {\em quasiparticle.})

If we take the second scenario Fig.~\ref{fig:qpqhfinal1}(b)
in the previous subsection,
the Majorana fermion $\eta_1$ associated with the
quasihole does not change sign, as the two
effects cancel out.
Consequently, the complex fermion $c=\eta_1 + i \eta_2 $
remains the same. The pair can now be annihilated.
On the other hand, in the first scenario Fig.~\ref{fig:qpqhfinal1}(a),
both the Majorana fermions $\eta_1, \eta_2$ change sign,
as well as the complex fermion: $c \rightarrow -c$.
Since $-c|0\rangle=0$, the pair can, again, still be annihilated.

Thus, if we impose an antiperiodic boundary condition in
$x$ direction, $\calT_x | \Psi_0 \rangle$ does give a groundstate
as the blocking effect is lifted.
Similarly, $\calT_y | \Psi_0 \rangle$ also gives a groundstate,
starting from the boundary condition $--$.

$\calT_y \calT_x | \Psi_0 \rangle$ does not, however, give
a groundstate.
Let us consider the process in sequence.
First we generate $\calT_x | \Psi_0 \rangle$, which is a groundstate.
It should be noted that, applying $\calT_x$ changes
the boundary condition along $y$ direction, as we have
discussed previously.
The boundary condition along $y$ direction
has changed from antiperiodic to periodic
in the new groundstate $\calT_x | \Psi_0 \rangle$.
Then we apply $\calT_y$ on the new groundstate with
the altered boundary condition;
the pair is now blocked from the annihilation
because of the same mechanism discussed in the previous
subsection.

Generalizing, we find that, among the $4(M+1)$ states
listed in~(\ref{eq:4M+4states}),
$M+1$ states of the form
\begin{equation}
 {\calT_x}^{2n+1} \calT_y | \Psi_0 \rangle ,
\end{equation}
for $n=0,1,2,\ldots,M$ do not exist as groundstates
owing to the leftover pair blocked from the annihilation.

As a result, starting from the boundary condition $--$,
we can generate $3(M+1)$ groundstates (including the original one).
They can be grouped into $3$ sectors, each containing $M+1$ states:
$M+1$ states of the form
\begin{equation}
 {\calT_x}^{2n} | \Psi_0 \rangle
\end{equation}
belongs to the boundary condition $--$,
\begin{equation}
 {\calT_x}^{2n} \calT_y | \Psi_0 \rangle
\end{equation}
to the boundary condition $+-$,
and
\begin{equation}
 {\calT_x}^{2n+1} | \Psi_0 \rangle
\end{equation}
to the boundary condition $-+$,
where $n=0, 1, 2, \ldots, M$.

This is again consistent with
the finding in Ref.~\cite{Read00}, obtained with
a quite different argument.
In Ref.~\cite{Read00}, the groundstates for
the boundary conditions $++$,$+-$, and $-+$
are argued to belong to the ``even electron number sector.''
While our argument does not tell the parity of the
electron number, it clarifies why
the groundstates are split into two sectors.

\subsection{Relation to Ising TQFT}

We now rephrase the preceding discussion in
the language of the Ising topological quantum field theory
(TQFT) and the $SU(2)_2$ Chern-Simons theory
with which it is closely related. The allowed topological charges
in the Ising TQFT are $1$, $\psi$, and $\sigma$,
with fusion rules: $\sigma\cdot\sigma \sim 1 + \psi$,
$\sigma\cdot\psi \sim \sigma$, $\psi\cdot\psi\sim 1$,
and anything fused with $1$ is itself. These topological
charges correspond to the vacuum, the neutral fermion,
and the half-flux quantum quasiparticle, respectively.
The different ground states on the torus correspond to the different
possible topological charges measured by a loop
encircling the torus around the $x$ direction. (The
different possible topological charges measured in
the $y$ direction gives a different basis for these states,
which is related to the other basis by the $S$-matrix.)
We will call these three ground states $\left|{1_x}\right\rangle$,
$\left|{\psi_x}\right\rangle$, $\left|{\sigma_x}\right\rangle$.
These correspond to the three ground states of the $M=0$
case. We will see in this language why there isn't a fourth one.

The action of $\calT_x$ and $\calT_y$ can be obtained by the logic
of sections \ref{sec:revOS} and \ref{sec:consistency}.
The action of $\calT_y$ on a state $\left| {a_x}\right\rangle$
with $a=1,\psi,\sigma$ is to increase the topological charge
$a$ by $\sigma$ according to the fusion rules:
\begin{equation}
{\calT_y} \left| {a_x}\right\rangle =  \left| \sigma\cdot a\: {_x}\right\rangle
\end{equation}
where $\sigma\cdot a$ is given by the fusion rules.
The action of $\calT_x$ on the states $\left|{1_x}\right\rangle$,
$\left|{\psi_x}\right\rangle$ is to multiply by the phase $\pm 1$
that results when a half-flux quantum quasiparticle is taken
around the vacuum or a neutral fermion, respectively.
\begin{equation}
{\calT_x} \left| {1_x}\right\rangle = \left| {1_x}\right\rangle\, ,
\hskip 0.65 cm
{\calT_x} \left| {\psi_x}\right\rangle = -\left| {\psi_x}\right\rangle
\end{equation}
In terms of these states, we can write a simultaneous eigenstate,
$\left|{\Psi_0}\right\rangle$, of $\calF \sim {\calT_y^2}$ and
$\calF_x \sim \calT_y^{-2}$ as:
\begin{equation}
\left|{\Psi_0}\right\rangle = \left(\left|{1_x}\right\rangle +
\left|{\psi_x}\right\rangle\right)/\sqrt{2}
\end{equation}
Then by applying $\calT_x$, $\calT_y$, we find:
\begin{eqnarray}
{\calT_x}\left|{\Psi_0}\right\rangle &=& \left(\left|{1_x}\right\rangle -
\left|{\psi_x}\right\rangle\right)/\sqrt{2}\cr
{\calT_y}\left|{\Psi_0}\right\rangle &=& \left|{\sigma_x}\right\rangle\cr
{\calT_y}{\calT_x}\left|{\Psi_0}\right\rangle &=&
{\calT_y}\left(\left|{1_x}\right\rangle -
\left|{\psi_x}\right\rangle\right)/\sqrt{2}
= \left(\left|{\sigma_x}\right\rangle -
\left|{\sigma_x}\right\rangle\right)/\sqrt{2} = 0
\end{eqnarray}
From the final line, we see that the naively-expected fourth state
vanishes, just as we saw in the previous section from
an analysis of the blocking mechanism. The above analysis
holds for $SU(2)_2$ Chern-Simons theory with
$1,\sigma,\psi$ replaced by the $j=0,\frac{1}{2},1$
representations of $SU(2)_2$.

\section{Higher Genus}

Now we extend our construction to general Riemann
surfaces with genus $g$.
As we have reviewed in Sec.~\ref{sec:revOS}, there are $g$ pairs
of intersecting elementary cycles
\begin{equation}
(\alpha_1, \beta_1), (\alpha_2, \beta_2), \ldots, (\alpha_g, \beta_g).
\end{equation}
As a commuting set of observables, we can use
\begin{equation}
 \calF_{\alpha_1}, {\calF_{\beta_1}}^{M+1},
 \calF_{\alpha_2}, {\calF_{\beta_2}}^{M+1}, \dots
 \calF_{\alpha_g}, {\calF_{\beta_g}}^{M+1}.
\end{equation}
The groundstates are labelled by their eigenvalues as
\begin{equation}
 | f_{\alpha_1}, f'_{\beta_1}, f_{\alpha_2}, f'_{\beta_2}, \ldots,
   f_{\alpha_g}, f'_{\beta_g} \rangle .
\end{equation}
Starting from the initial state
\begin{equation}
|\Psi^{(0)}\rangle =
 | {f_{\alpha_1}}^{(0)}, {f'_{\beta_1}}^{(0)},
   {f_{\alpha_2}}^{(0)}, {f'_{\beta_2}}^{(0)}, \ldots,
   {f_{\alpha_g}}^{(0)}, {f'_{\beta_g}}^{(0)} \rangle ,
\end{equation}
we can construct various states by applying $\calT$ operators.
For each pair $(\alpha_j, \beta_j)$,
by applications of $\calT_{\alpha_j}$ and $\calT_{\beta_j}$,
we can generate $4(M+1)$ set of different eigenvalues
\begin{eqnarray}
f_{\alpha_j} &=&     e^{-i \pi m_j/(M+1)} {f_{\alpha_j}}^{(0)},
\label{eq:faj}
\\
f'_{\beta_j} &=&     \tau_j {f'_{\beta_j}}^{(0)},
\label{eq:fbj}
\end{eqnarray}
where $m_j=0,1,2,\ldots,2M+1$ and $\tau_j =\pm 1$.
In total, we can generate $[4(M+1)]^g$ different set of
eigenvalues.

However, not all of them actually give groundstates,
because of the blocking effect discussed in the previous subsection.
Let us start from antiperiodic boundary conditions
along all cycles: $---- \ldots --$.
Among the states with
$4(M+1)$ different eigenvalues $f_{\alpha_j}, f'_{\beta_j}$,
as discussed on the torus, $3(M+1)$ of them
($m_j$ even or $\tau_j=1$) immediately
give groundstates because the quasiparticle-hole pair can be
annihilated. The other $M+1$ ($m_j$ odd and $\tau_j=-1$)
do not directly give groundstates
owing to the leftover pair which are blocked from annihilation.

By choosing one of the $3(M+1)$ set of eigenvalues from
each pair of elementary cycles $(\alpha_j, \beta_j)$,
we can readily obtain $[3(M+1)]^g$ groundstates without
any leftover pair.
However, they do not exhaust all the groundstates
which can be obtained in the present construction.

Suppose we have two leftover quasiparticle-hole pairs which are
blocked from annihilation, after the applications of $\calT$
operators.
Let us denote the Majorana fermion boundstates
just after the creation of the pairs,
as $\eta^{(k)}_l$,
where $k=1,2$ is the index for pairs, and $l=1$ and $l=2$
refer to the quasiparticle and quasihole, respectively.
After the two $\calT$ processes, suppose that
the Majorana fermions carried by the quasiholes
$\eta^{(1)}_2$ and $\eta^{(2)}_2$ flip their signs.
In other words, at this point, the Majorana fermions are given by
\begin{equation}
  \eta^{(1)}_1,
  - \eta^{(1)}_2,
  \eta^{(2)}_1,
  - \eta^{(1)}_2 .
\end{equation}
This change in internal state blocks the annihilation
of the pairs, as discussed around eq.~(\ref{eq:psi2dag}).
However, here we can exchange the two quasiholes twice.
This induces a change in the Majorana fermions as
\begin{equation}
  (- \eta^{(1)}_2, - \eta^{(2)}_2) \rightarrow
  (- \eta^{(2)}_2, + \eta^{(1)}_2) \rightarrow
  ( + \eta^{(1)}_2,+ \eta^{(2)}_2) .
\end{equation}
Thus the phase acquired during $\calT$ processes
can be cancelled by the exchange of two quasiparticles.
After this ``mending'' process, we can annihilate both of
the leftover quasiparticle-quasihole pairs.

When there is an
even number of leftover pairs after the application
of $\calT$ operators, we can actually annihilate all of
them with the mending procedure described above.
Among the $[4(M+1)]^g$ states labelled by different
eigenvalues~(\ref{eq:faj}) and~(\ref{eq:fbj}),
the number of states with $k$ leftover pairs is given by
\begin{equation}
 \left( \begin{array}{c} g \\ k \end{array} \right)
 \; [3(M+1)]^{g-k} (M+1)^k
 = \frac{g!}{k! (g-k)!} [3(M+1)]^{g-k} (M+1)^k .
\end{equation}
Thus, the total number of the groundstates we can obtain with
the mending procedure is given, using binomial theorem, by:
\begin{eqnarray}
N_g &=& \sum_{k=0}^g \frac{1+(-1)^k}{2}
\frac{g!}{k! (g-k)!} [3(M+1)]^{g-k} (M+1)^k \nonumber \\
\nonumber \\
&=& (M+1)^g 2^{g-1}(2^g+1).
\end{eqnarray}
This exactly agrees with the result~(\ref{eq:evenNg})
by Read and Green~\cite{Read00}
for the ``even electron number sector''.

We note that, as in the case of torus,
application of a $\calT$ operator changes the boundary
condition of the system.
There are certain boundary conditions, for which the groundstate
is not accessible by application of $\calT$ operators
on the initial groundstate with the $---- \ldots ---$ boundary
condition.
The inaccessible groundstates in these boundary conditions
belong to the other sector which is separate from the ones
discussed above.
In the context of Ref.~\cite{Read00}, this would
correspond to the ``odd electron number sector''.

One of such boundary conditions is
$++---- \ldots ---$, namely the periodic boundary conditions
on the first pair of elementary cycles ($\alpha_1, \beta_1$)
and antiperiodic boundary conditions on all the other elementary
cycles.
Let us start from a groundstate in this boundary condition.
Acting on the first pair of elementary cycles,
applying $\calT_{\alpha_1}$, $\calT_{\beta_1}$,
or $\calT_{\alpha_1} \calT_{\beta_1}$ produces a leftover
pair which cannot be annihilated by itself.
In this case, one must produce an odd number of leftover pairs
from the other elementary cycles in order to obtain a
groundstate with the mending processes.

On the other hand, if we do not apply any of
the $\calT$ operators on the first pair of elementary
cycles ($\alpha_1, \beta_1$), we would have no
leftover pair from these cycles.
In this case, we must produce an even number (of course
including zero) of leftover
pairs from the other elementary cycles.

In either case, we can apply ${\calT_{\alpha_j}}^{2n_j}$
with $n_j=0,1,2,\ldots,M$
on each elementary cycles $\alpha_j$,
to produce distinct groundstates,
without changing the boundary conditions and without
producing any leftover pair.

Thus the total number of the groundstates reads
\begin{eqnarray}
N_g &=& (M+1)^g
 \big[
 \big( 1 \times \sum_{k=0}^{g-1} \frac{1+(-1)^k}{2}
 \left( \begin{array}{c} g-1 \\ k \end{array} \right)
 \; 3^{g-1-k} 1^k
 \big)
\nonumber \\
 && +
 \big( 3 \times \sum_{k=0}^{g-1} \frac{1-(-1)^k}{2}
 \left( \begin{array}{c} g-1 \\ k \end{array} \right)
\; 3^{g-1-k} 1^k
 \big) \big]
\nonumber \\
&=& (M+1)^g 2^{g-1} (2^g -1 ) .
\end{eqnarray}
This is again in exact agreement with the result in
Ref.~\cite{Read00} on the odd electron number sector.

The total number of states in the two sectors is $[4(M+1)]^g$,
which is equal to what the number of the groundstates would
be in the absence of the blocking mechanism.
This implies that there are only two sectors (``even electron
number'' and ``odd electron number'') even for higher genus $g$.
We should remember, however, that the present construction
does not necessarily give the exact number of groundstates,
although it would give a lower bound.
In general, there can be more groundstates due to other
mechanisms not covered in the present argument.
In the case of the Pfaffian states, it appears that
the present construction does give all the degenerate
groundstates, considering the agreement with Ref.~\cite{Read00}.

\section{Topological degeneracy of $p+ip$ superconductors}
\label{sec:p+ip}

It follows from \cite{Read00}, and has been explicitly verified by
Ivanov~\cite{Ivanov01} and Stern \emph{et.~al.}~\cite{Stern04} that
a $p+ip$-wave superconductor in its weak-pairing phase in 2 dimensions
exhibits the non-Abelian statistics identical (up to the Abelian phase) to
that in the Pfaffian states.
It is then natural to ask what is the topological degeneracy
of the $p+ip$ superconductor.
It turns out that, although the non-Abelian statistics
is the same, the framework developed in the previous sections
for the Pfaffian states does not directly apply to
the $p+ip$ superconductor.
Nevertheless, a slightly different set of operations
can be used to construct the degenerate groundstates.

Before moving on to the $p+ip$ superconductor with the non-Abelian
statistics, let us discuss the simpler $s$-wave superconductor.
Although the $s$-wave superconductor has been known for a long time,
it was only recently that its topological order was emphasized by
Hansson, Oganesyan and Sondhi~\cite{SuperTopo}. Here we review the
topological degeneracy of ``ordinary'' $s$-wave superconductor as
discussed in Ref.~\cite{SuperTopo}. In order to draw parallel with
the previous sections, we try to keep the discussion as elementary
as possible, without relying on the effective field theory.

The superconductor should be gapless because of the Nambu-Goldstone
mechanism, if one does not include the dynamics of the electromagnetic
gauge field.
However, once the gauge field dynamics (or equivalently, the long-range
Coulomb interaction) is taken into account, the system has a finite
gap. This is the Anderson-Higgs mechanism.
Here we consider such a superconducting system with a non-vanishing
excitation gap above the groundstate.
For simplicity, let us first consider the torus.
As elementary excitations, the superconductor
has Bogoliubov quasiparticle/hole and vortex/antivortex
with half flux quantum $\Phi_0/2$.
The difference in the superconductor compared to the
Quantum Hall case is that the
quasiparticle/hole are different from the
vortex/antivortex.

We introduce the processes similar to $\calT_{x,y}$ used
in the Pfaffian Quantum Hall states, separately
for these objects.
Namely,
$\calA_{\alpha}$ ($\alpha=x,y$)
is defined as a creation of quasiparticle/hole,
followed by the quasiparticle encircling the torus in
$\alpha$ direction and then the pair annihilation.
Similarly, $\calB_{\alpha}$  ($\alpha=x,y$)
is a creation of vortex/antivortex, vortex encircling
the torus in $\alpha$ direction and then
the pair annihilation.

As it was discussed in the Quantum Hall case,
$\calB_{\alpha}$ introduces a half flux quantum $\Phi_0/2$
into a ``hole'' of the torus.
In the other words, $\calB_x$ and $\calB_y$ twist the boundary
condition (periodic to antiperiodic or antiperiodic to periodic)
in $y$ and $x$ directions, respectively.

On the other hand, the Bogoliubov quasiparticle/hole
has no definite charge.
Nevertheless, it has a definite charge $e$ modulo $2e$,
as the quasiparticle creation operator is a linear
combination of electron creation/annihilation operator.
The charge of an elementary excitation
is not definite because of the condensate
of Cooper pairs, but it is nevertheless well-defined modulo $2e$
because the condensate can only absorb/emit
the Cooper pair with charge $2e$.
Because of this ``odd charge'', the Aharonov-Bohm phase
gained by the quasiparticle during the encircling process
is changed by $\pi$ due to the insertion of half flux quantum.

Thus we obtain the anticommutation relations
\begin{eqnarray}
\calA_x \calB_y &=& - \calB_y \calA_x,
\label{eq:axby}
 \\
\calA_y \calB_x &=& - \calB_x \calA_y.
\label{eq:aybx}
\end{eqnarray}

Unlike in the Quantum Hall case,
the quasiparticle here does not carry magnetic flux.
Thus we have the commutation relation
\begin{equation}
\big[ \calA_x , \calA_y \big] = 0 .
\label{eq:axay}
\end{equation}
This may also be understood
as a consequence~\cite{OshikawaSenthil} of the fact that
the quasiparticle is a fermion.

Thus the groundstate on a torus may be labelled by
the eigenvalues of $\calA_x$ and $\calA_y$
which are simultaneously diagonalizable.
Let
\begin{eqnarray}
\calA_x | a_x , a_y \rangle &=& a_x | a_x , a_y \rangle, \\
\calA_y | a_x , a_y \rangle &=& a_y | a_x , a_y \rangle.
\end{eqnarray}
The anticommutation relations~(\ref{eq:axby}) and
(\ref{eq:aybx}) implies that, starting from a groundstate
$\inita$, we can generate $3$ other
groundstates as follows:
\begin{eqnarray}
\calB_y \inita &\sim&  | -a_x^{(0)}, a_y^{(0)} \rangle, \\
\calB_x \inita &\sim&  | a_x^{(0)}, - a_y^{(0)} \rangle, \\
\calB_x \calB_y \inita &\sim&  | - a_x^{(0)}, - a_y^{(0)} \rangle .
\end{eqnarray}
In total, there are $4$ groundstates on a torus.

It is straightforward to generalize the above argument to
a general Riemann surface with genus $g$.
There are are $g$ pairs of intersecting elementary cycles.
The $\calA$ operators defined on every elementary
cycle commute with each other, and are simultaneously
diagonalizable.
Operating $\calB_{\gamma'}$ flips the sign of the eigenvalue
of $\calA_\gamma$, where ${\gamma'}$ and $\gamma$ are
intersecting elementary cycles.
Thus each eigenvalue of $\calA_\gamma$ can be flipped, and
there are $2^{2g}$ degenerate groundstates.

Now let us move on to the topological degeneracy of
a $p+ip$ superconductor.
The nature of quasiparticle is the same as in the $s$-wave case.
The new aspect here is the Majorana fermion boundstate
at a vortex/antivortex.
It leads to the non-Abelian statistics of vortex/antivortex
as elucidated by Ivanov.

The commutativity of $\calA$ operators~(\ref{eq:axay}) and
anticommutation relations~(\ref{eq:axby}),(\ref{eq:aybx})
between $\calA$ and $\calB$ operators remain the same.
However, as in the Quantum Hall case, the Majorana
fermion boundstate of a vortex/antivortex feels
the effect of the magnetic flux.
Starting from the antiperiodic boundary conditions
on the both $x$ and $y$ directions,
either $\calB_x$ or $\calB_y$ gives a new groundstate
as it flips the eigenvalue of $\calA_y$ or $\calA_x$,
respectively.
Here the vortex/antivortex can be pair annihilated,
as the internal state of the Majorana fermions after
the vortex encircling the torus is the same
as just after the pair creation.
However, once $\calB_x$ is applied, the boundary
condition on $y$ direction is changed to the
periodic one.
Then applying $\calB_y$ no longer gives a new groundstate,
as the vortex/antivortex pair is blocked from the
pair annihilation, because of the change in the
internal state as discussed on the Pfaffian Quantum Hall states.

Therefore, on the torus, we can generate only $3$ groundstates
\begin{equation}
\inita, \;\;\; \calB_x \inita, \;\;\; \calB_y \inita .
\end{equation}
The number of the degenerate groundstates turns out to
be identical to that for $M=0$ bosonic Pfaffian state,
although for a slightly different mechanism.
The role played by $\calF$ operators in the Pfaffian states
are provided by $\calA$ operators instead.
Except for this difference, the analysis is very much
in parallel.

Generalization to general genus $g$ can be also done
in a similar way to the discussion on the Pfaffian states.
One can produce $2^{2g}$ candidate groundstates
by applying $\calB$ operators and flipping the signs
of the eigenvalues of $\calA$ operators.
However, if $\calB_\gamma$ and $\calB_{\gamma'}$ are
applied for an intersecting pair $\gamma,{\gamma'}$,
a vortex/antivortex pair is left, being blocked from
annihilation.
Nevertheless, if the number of such leftover pairs is
even, we can eventually annihilate all of them
using the mending process.
Thus the number of degenerate groundstates
is again exactly the same as that of
the $M=0$ bosonic Pfaffian state.

Finally, we remark on the difference between the weak-pairing
(non-Abelian) and strong-pairing (Abelian) phases of a $p+ip$
superconductor. The main difference between these two cases (from
the statistics viewpoint) is that the zero-energy Majorana mode in a
vortex core is absent in the strong-pairing phase
\cite{Read00,TDNZZ}. Instead, there are two boundstates whose
energies are split around 0. Hence the ``elementary'' excitations in
this case are (empty) vortices and Bogoliubov quasiparticles, just
as it would be for a generic $s$-wave superconductor discussed
earlier. (The quasiparticle/quasihole states localized in the vortex
cores are just the dyonic boundstates in this language.) Hence the
reasoning presented for the $s$-wave case applies here and we end up
with the total of $2^{2g}$ degenerate groundstates, effectively
combining both disjoint sectors present in the weak--pairing phase.

\section{Summary}

In this paper,
we extended the idea in Ref.~\cite{OshikawaSenthil} to
a system with quasiparticles obeying non-Abelian statistics,
and analyzed the Moore-Read Pfaffian state~\cite{Moore91}
in particular, based on the physical
pictures given in Refs.~\cite{Read00,Ivanov01}.

We gave an explicit construction of distinct groundstates using physical
processes, and derived the groundstate degeneracy on Riemann surfaces
with general genus $g$.
This exactly reproduced the groundstate degeneracy given
by Read and Green~\cite{Read00} based on the correspondence
to ``spin structures'' in differential geometry.

We also found that the quasiparticle charge
imposes certain requirements on the non-Abelian statistics
of the quasiparticles. In the case of the Pfaffian state,
four consecutive exchanges of quasiparticles reduces
to an Abelian factor~\cite{Ivanov01}, as is required from the charge
argument.
Moreover, the charge of the quasiparticle
also restricts the ``Abelian'' part of the statistics.

We also discussed the topological degeneracy of the
$p+ip$ superconductor, which is closely related to
the Pfaffian states.
Although it requires a somewhat different construction
based on Ref.~\cite{SuperTopo},
the topological degeneracy turned out to be identical
to that of the $M=0$ bosonic Pfaffian state.

Our argument and observations would be applicable
to other non-Abelian systems.
We hope that the present method
help understanding of the nature of non-Abelian states,
and future developments on the subject.

\section*{Acknowledgements}

The authors thank Akira Furusaki and T. Senthil
for stimulating discussions.
The present work was done partially
in the 2005 Summer Workshop
``Gauge Theories and Fractionalization in Correlated Quantum Matter'',
at the Aspen Center for Physics,
and in the 2006 Program ``Topological Phases and Quantum Computation'' at the Kavli Institute for Theoretical Physics,
UC Santa Barbara,
supported by the National Science Foundation under
Grant No. PHY99-07949.
M.O. is supported in part by
a 21st Century COE Program at Tokyo Institute of Technology
``Nanometer-Scale Quantum Physics'' from MEXT of Japan,
while he belonged to Department of Physics, Tokyo Institute
of Technology until March 2006.
Y.B.K. is supported by the NSERC of Canada, the CRC program,
the Canadian Institute for Advanced Research, and
KRF-2005-070-C00044. C.N. is supported by
the National Science Foundation under grant
DMR-0411800 and by the Army Research Office
under grant W911NF-04-1-0236.


\end{document}